\documentclass[conference]{IEEEtran}
\usepackage{cite}
\usepackage{amsmath,amssymb,amsfonts}
\usepackage{graphicx}
\usepackage{textcomp}
\usepackage{xcolor}
\usepackage{authblk}

\usepackage{booktabs}
\usepackage{epsfig}
\usepackage{latexsym}
\usepackage{multirow}
\usepackage{stfloats}
\usepackage{epstopdf}
\usepackage{color}  
\usepackage{tabularx} 
\usepackage{enumerate}
\usepackage{array}
\graphicspath{{./Figures/}}
\usepackage{color}
\usepackage{bbm}
\usepackage{caption}
\usepackage{bm}
\usepackage[tight,footnotesize]{subfigure}
\usepackage{balance}
\usepackage{mathrsfs}
\usepackage{verbatim}
\allowdisplaybreaks[4]
\usepackage{dsfont}
\usepackage{verbatim}
\usepackage{tikz}
\usepackage{setspace}
\usepackage{diagbox}
\usepackage[framemethod=tikz]{mdframed}
\usepackage{multicol}
\usepackage{environ}
\usepackage{tikz}
\usepackage{stfloats}
\usepackage{algpseudocode}
\usepackage{graphics}
\usepackage{epsfig}
\usepackage{amsthm}
\usepackage{authblk}

\ifCLASSOPTIONcompsoc
\usepackage[caption=false, font=normalsize, labelfont=sf, textfont=sf]{subfig}
\else
\usepackage[caption=false, font=footnotesize]{subfig}

\usepackage{bm}
\def\BibTeX{{\rm B\kern-.05em{\sc i\kern-.025em b}\kern-.08em
    T\kern-.1667em\lower.7ex\hbox{E}\kern-.125emX}}
\begin{document}

\title{Terahertz Channel Measurement and Analysis on a University Campus Street}

\author{Yiqin Wang}
\author{Yuanbo Li}
\author[2]{Yi Chen}
\author[2]{Ziming Yu}
\author{Chong Han}
\affil{Terahertz Wireless Communications (TWC) Laboratory, Shanghai Jiao Tong University, China.\authorcr Email: \{wangyiqin, yuanbo.li, chong.han\}@sjtu.edu.cn}
\affil[2]{Huawei Technologies Co., Ltd., China. Email: \{chenyi171, yuziming\}@huawei.com}

%\author{\IEEEauthorblockN{Yiqin Wang}
%\IEEEauthorblockA{\textit{University of Michigan-Shanghai}\\
%\textit{Jiao Tong University Joint Institute}\\
%\textit{Shanghai Jiao Tong University}, Shanghai, China\\
%Email: wangyiqin@sjtu.edu.cn}
%\and
%\IEEEauthorblockN{Chong Han}
%\IEEEauthorblockA{\textit{University of Michigan-Shanghai}\\
%\textit{Jiao Tong University Joint Institute}\\
%\textit{Shanghai Jiao Tong University}, Shanghai, China\\
%Email: chong.han@sjtu.edu.cn}}

\maketitle
\begin{abstract}
Owning abundant bandwidth resource, the Terahertz (0.1-10~THz) band is a promising spectrum to support sixth-generation (6G) and beyond communications. As the foundation of channel study in the spectrum, channel measurement is ongoing in covering representative 6G communication scenarios and promising THz frequency bands. In this paper, a wideband channel measurement in an L-shaped university campus street is conducted at 306-321~GHz and 356-371~GHz.
%\rev{The sounder system consists of a vector network analyzer (VNA)-based channel sounder, and a directional antenna equipped at the receiver to resolve multi-path components (MPCs) in the angular domain.}
In particular, ten line-of-sight (LoS) and eight non-line-of-sight (NLoS) points are measured at the two frequency bands, respectively. In total, 6480 channel impulse responses (CIRs) are obtained from the measurement, based on which multi-path propagation in the L-shaped roadway in the THz band is elaborated to identify major scatterers of walls, vehicles, etc. in the environment and their impact on multi-path components (MPCs). Furthermore, outdoor THz channel characteristics in the two frequency bands are analyzed, including path losses, shadow fading, cluster parameters, delay spread and angular spread. In contrast with the counterparts in the similar outdoor scenario at lower frequencies, the results verify the sparsity of MPCs at THz frequencies and indicate smaller power spreads in both temporal and spatial domains in the THz band.
\end{abstract} 
\section{Introduction}

As a promising spectrum band for sixth-generation (6G) communications, the Terahertz (0.1-10~THz) band is anticipated to meet the demand of wireless systems for data rates to reach terabits per second (Tbps)~\cite{akyildiz2022terahertz,chen2021terahertz,rappaport2019wireless}.
In this new spectrum, the design of wireless systems relies on the study of THz wireless channels, which includes physical measurement of channels, analysis of channel characteristics, among others~\cite{channel_tutorial}.
Compared with well investigated centimeter-wave (cmWave) and millimeter-wave (mmWave) bands, the exploration of THz wireless channels, especially the physical measurement of THz wireless channels, is far from sufficient to cover all representative 6G communication scenarios and promising THz frequencies.

Channel measurement campaigns have been conducted mainly at 140~GHz, 190~GHz, and 220~GHz~\cite{channel_tutorial,abbasi2022thz,ju2022sub}, i.e., in the ``sub-THz'' band that overlaps with the mmWave band~\cite{rappaport2019wireless}.
By contrast, wireless channel measurements above 300~GHz are still inadequate.
In the up-to-date acts of the World Radiocommunication Conference 2019 (WRC-19), the frequency bands from 275~GHz to 450~GHz have identified for the implementation of fixed and land mobile service applications~\cite{WRC-19-FINAL-ACTS}.
Some existing measurement studies in these frequency bands are conducted in indoor hallways~\cite{wang2022thz, li2022channel}, an aircraft carbin~\cite{doeker2022channel}, a courtyard~\cite{undi2021angle}, and railway and vehicular communication scenarios~\cite{guan2021channel,eckhardt2021channel}.

Nevertheless, channel measurement campaigns above 300~GHz hardly go to outdoor scenarios. Besides, most of these studies use correlation-based channel sounder with measuring bandwidth no more than 8~GHz.
To fill this research gap, the contributions of this work include that we present a wideband vector network analyzer (VNA)-based channel measurement campaign conducted in a typical outdoor scenario in two frequency bands above 300~GHz, with measuring bandwidth of 15~GHz.
In specific, the measurement campaign is conducted in an L-shaped street on a university campus at 306-321~GHz and 356-371~GHz. The scenario features one street along the parking area, and the other perpendicular street canyon between two buildings.
%The sounder system consists of a VNA-based sounder and a directional antenna to resolve multi-path components (MPCs) in the angular domain.
Both line-of-sight (LoS) and non-line-of-sight (NLoS) cases are investigated, with 18 measuring positions and 6480 channel impulse responses (CIRs) in total. In light of the measurement results, we retrive the propagation of multi-path components (MPCs) in the L-shaped street. Major MPC contributors in the environment, namely walls, vehicles, traffic signs and lamp poles, are identified, and their impacts by means of transmission, reflection, and scattering of MPCs are explored. Channel characteristics in the two frequency bands are analyzed in depth, including the path losses, shadow fading, cluster parameters, delay spread and angular spreads. Furthermore, we compare our results with those in the Urban Micro (UMi)-street canyon scenario specified in 3rd Generation Partnership Project (3GPP) Technical Report (TR)~38.901~\cite{3gpp38901}, which indicates the sparsity of MPCs and smaller power spreads in both temporal and spatial domains in the THz band.

The remainder of this paper is organized as follows.
The wideband VNA-based channel sounder system and the measurement campaign are introduced in Sec.~\ref{section: campaign}. We then present channel measurement results in Sec.~\ref{section: results}. Specifically, major scatterers in the environment and their impact on MPCs are investigated, and channel characteristics are analyzed.
Finally, the paper is concluded in Sec.~\ref{section: conclusion}.
\section{Channel Measurement Campaign} \label{section: campaign}

In this section, we introduce our VNA-based channel measurement platform, and describe the THz measurement campaign at 306-321~GHz and 356-371~GHz in a L-shaped street on the campus of Shanghai Jiao Tong University (SJTU). 

\subsection{Channel Measurement System}
Our channel measurement platform supports the frequency ranging from 260~GHz to 400~GHz. As detailed in~\cite{wang2022thz}, the system is composed of THz transmitter (Tx) and receiver (Rx) modules and the Ceyear 3672C VNA. The system calibration procedure is described in detail in our previous works~\cite{wang2022thz, li2022channel}.
\begin{comment}
The VNA generates radio frequency (RF) and local oscillator (LO) sources.
%The RF signal is multiplied by 27 to reach the carrier frequencies ranging from 260~GHz to 400~GHz. The LO signal is multiplied by 24 to reach the frequencies ranging from 260.0076~GHz to 400.0076~GHz. As a result, the mixed intermediate frequency (IF) signal has the frequency of 7.6~MHz.
The signals from RF and LO sources are multiplied by 27 and 24, respectively. As a result, the transmitted signal reaches the carrier frequencies ranging from 260~GHz to 400~GHz, and the mixed intermediate frequency (IF) signal has the frequency of 7.6~MHz.
%and therefore is free of the error caused by the cables when they carry high-frequency signals.
The IF signals at Tx and Rx modules are both sent back to the VNA, and the transfer function of the channel is calculated as the ratio of the two frequency responses. Interestingly, power of the sounder systems is supplied by a hybrid vehicle parking in ....
\end{comment}
In this measurement campaign, we investigate two frequency bands, 306-321~GHz and 356-371 GHz, which both cover the bandwidth of 15~GHz. The frequency sweeping interval is 2.5~MHz, resulting in 6001 sweeping points for each pair of Tx-Rx directions. Therefore, the space resolution is 2~cm and the maximum detectable path length is 120~m.
To compensate for the high path loss in the THz band, waveguides and WR2.8 horn antenna are installed, with gains around 7~dBi (3-dB beamwidth around 60$^\circ$) at the Tx side and 25-26~dBi (3-dB beamwidth around 8$^\circ$) at the Rx side, respectively.

Regarding the mechanical part of the measurement system, the Tx and Rx are lifted to reach the height of 3~m and 1.75~m, respectively. The Tx is static, while by contrast, to receive MPCs, the Rx is driven by the rotator to scan the complete azimuth domain and from $-20^\circ$ to $20^\circ$ in the elevation domain, with the step of $10^\circ$. Key parameters of the measurement are summarized in Table~\ref{tab:system_parameter}.

% \begin{figure}
%     \centering
%     \subfigure[The RF part.]{
%     \includegraphics[width=0.9\linewidth]{figures/RF_system.png}
%     }
%     \\
%     \subfigure[The mechanical part.]{
%     \includegraphics[width=0.6\linewidth]{figures/platform.png}
%     }
%     \caption{The 260-400~GHz channel measurement system.}
%     \label{fig:platform}
% \end{figure}

\begin{table}
  \centering
  \caption{Parameters of the measurement.}
    \begin{tabular}{ll}
    \toprule
    Parameter & Value \\
    \midrule
    Frequency band              & 306-321, 356-371~GHz \\
    Bandwidth                   & 15~GHz \\
    Sweeping interval           & 2.5~MHz \\
    Sweeping points             & 6001 \\
    Time resolution             & 66.7~ps \\
    Space resolution            & 2~cm \\
    Maximum excess delay        & 400~ns \\
    Maximum path length         & 120~m \\
    Tx antenna gain             & 7~dBi \\
    Tx 3-dB beamwidth           & 60$^\circ$ \\
    Rx antenna gain             & 25-26~dBi \\
    Rx 3-dB beamwidth           & 8$^\circ$ \\
    Tx height          & 3~m \\
    Rx height          & 1.75~m \\
    Rx azimuth angles   & $[0^\circ:10^\circ:360^\circ]$ \\
    Rx elevation angles & $[-20^\circ:10^\circ:20^\circ]$ \\
    Average noise floor         & -180~dBm \\
    Dynamic range               & 119~dB \\
    \bottomrule
    \end{tabular}
  \label{tab:system_parameter}
\end{table}

\subsection{Measurement Deployment}
\begin{figure*}
    \centering
    \subfigure[The top view of the measurement campaign.]{
    \includegraphics[width=0.46\linewidth]{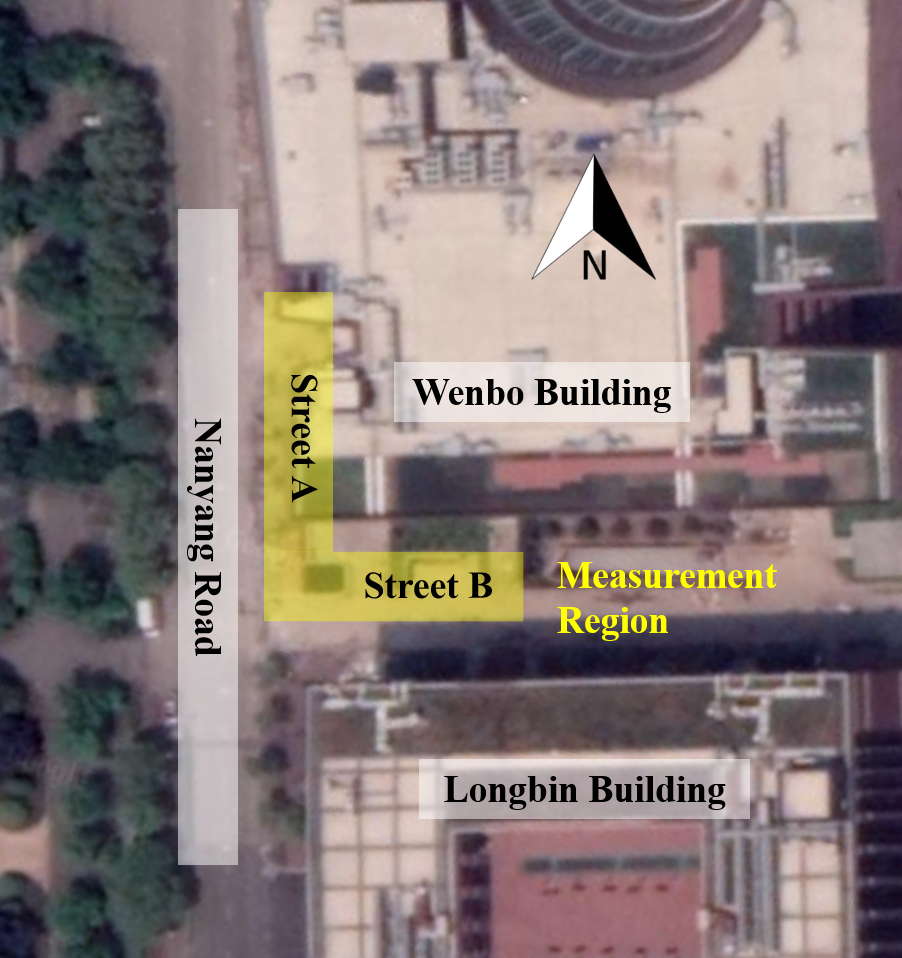}
    }
    \begin{minipage}[b]{0.5\linewidth}
    \centering
    \subfigure[The photo of the environment.]{
    \includegraphics[width=0.95\linewidth]{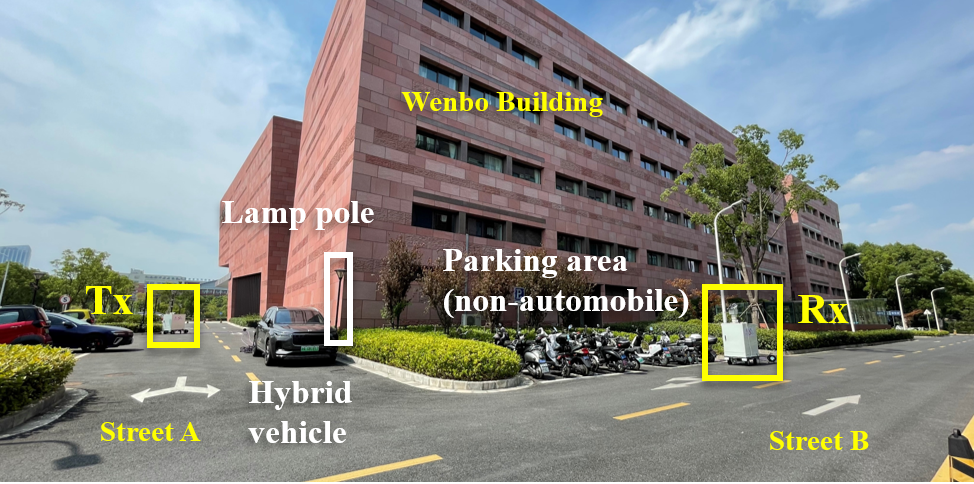}
    }\\
    \subfigure[The photo of the west street (Street~A).]{
    \includegraphics[width=0.95\linewidth]{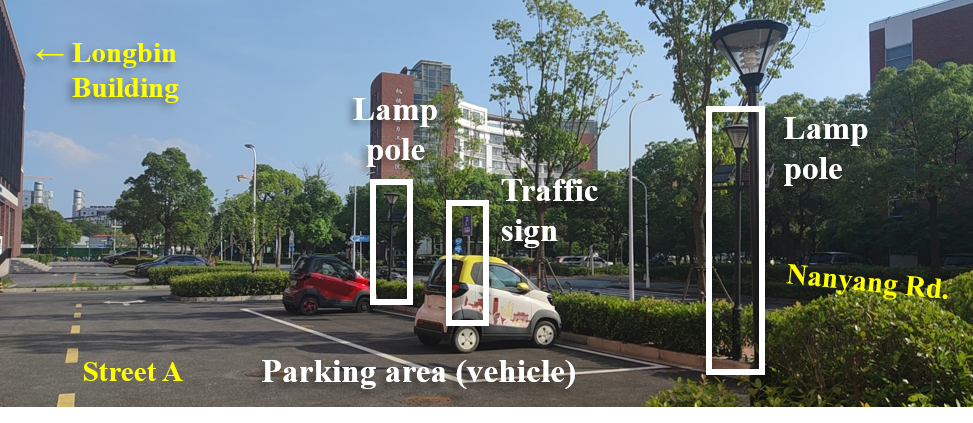}
    }
    \end{minipage}
    \caption{Measurement campaign in the L-shaped street on the SJTU campus.}
    \label{fig:environment}
\end{figure*}

As shown in Fig.~\ref{fig:environment}, the measurement campaign is conducted in the outdoor street bounded by Wenbo Building, Longbin Building, and Nanyang Road on the SJTU campus. Specifically, we focus on the L-shaped turning at the crossroad, depicted as the yellow region, with buildings, vehicles, pedestrians, plants, etc. The west street (Street~A) is parallel to Nanyang Road, and is between the west wall of Wenbo Building and a parking area. Vehicles, lamp poles and traffic signs are found along Street~A, and one lamp pole is at the southwest corner of Wenbo Building. The other street canyon (Street~B) between two buildings is about 22.32~m wide. Plants and a non-automobile parking area are located in the south of Wenbo Building along Street~B.

\begin{figure}
    \centering
    \includegraphics[width=0.95\linewidth]{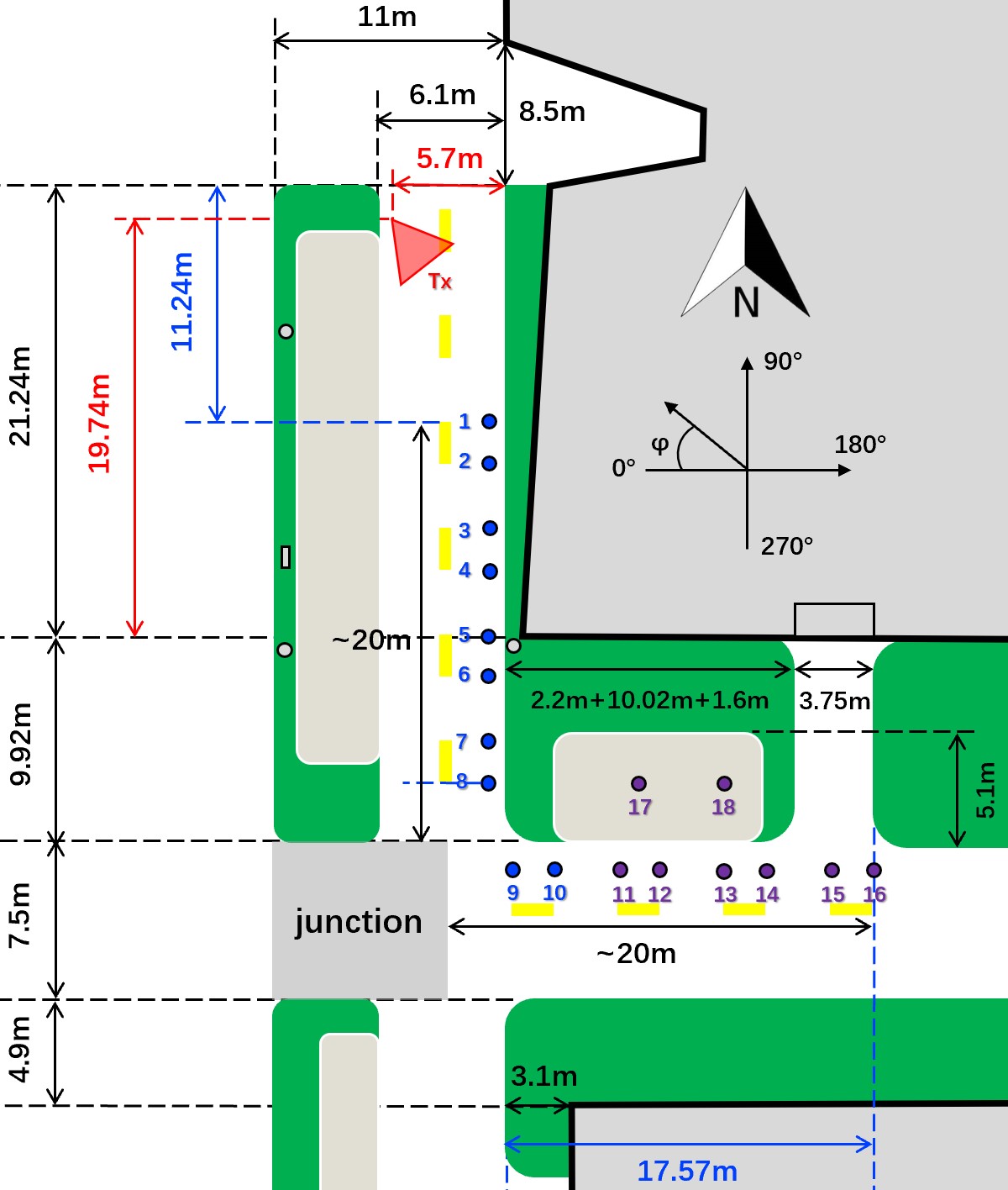}
    \caption{Measurement deployment in the L-shaped street on the SJTU campus (Tx, red; LoS Rx, blue; NLoS Rx, violet).}
    \label{fig:deployment}
\end{figure}

The measurement deployment is illustrated in Fig.~\ref{fig:deployment}. We set west when the horizontal angle $\varphi$ equals to 0$^\circ$, while the value of $\varphi$ increases clockwise. Tx is located at the northwest corner in Street~A, i.e., 29.66~m away from the junction of two streets. The directivity of Tx points to the southeast direction, with the fixed horizontal angle of $\varphi$ = 230$^\circ$ and elevation angle of $\theta$ = 0$^\circ$. The positions of Rx are distributed along two segments of the L-shaped street, labeled as Rx1-Rx8 and Rx9-Rx16. The farthest point is about 20~m away from the junction of two streets. Moreover, two additional Rx points, Rx17 and Rx18, are located in the non-automobile parking area, to simulate user equipment in the shadow of a building. To classify, Rx1 to Rx10 belong to the LoS case, while Rx11 to Rx18 belong to the NLoS case.

Interestingly, the source power of the sounder systems is supplied by a hybrid vehicle, which is parked in the parking area in Street~A in the LoS case. By contrast, due to the limit of cable length, it is parked at the southwest corner of Wenbo Building when we measure channel responses at NLoS points, as shown in Fig.~\ref{fig:environment}(b). Note that this vehicle creates transmission, reflection and scattering in the measurement.


%\subsection{\rev{Calibration and Data Post-processing}}

\section{Channel Analysis and Characterization} \label{section: results}
In this section, power-delay-angle profile (PDAP) is analyzed for multi-path characterization in the L-shaped street. Moreover, channel features in two frequency bands are analyzed in depth, including path losses, shadow fading, cluster parameters, delay spread and angular spread.

\subsection{Power-Delay-Angle-Profile (PDAP) Analysis} \label{sec:MPC_analysis}

\begin{figure*}
    \centering
    \subfigure[PDAP at Rx1-4.]{
    \includegraphics[width=0.75\linewidth]{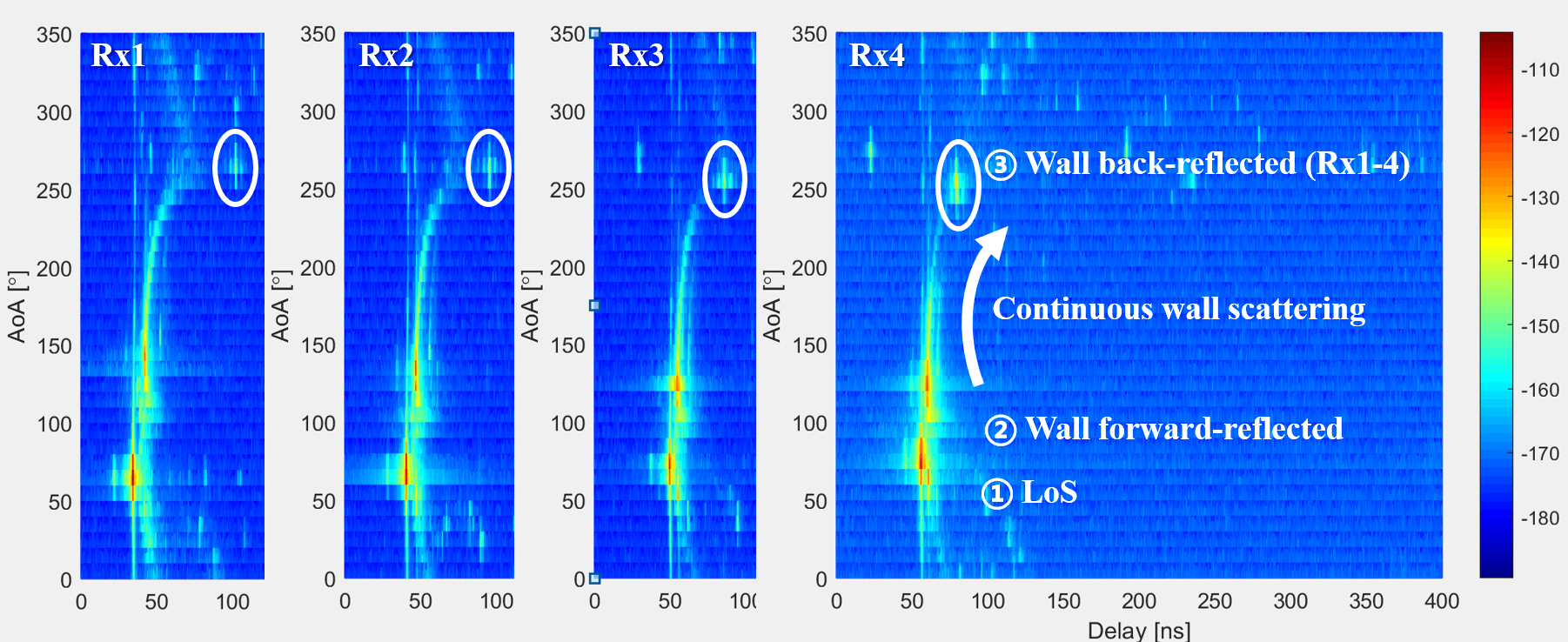}
    }
    \subfigure[Multi-path at Rx1-4.]{
    \includegraphics[width=0.2\linewidth]{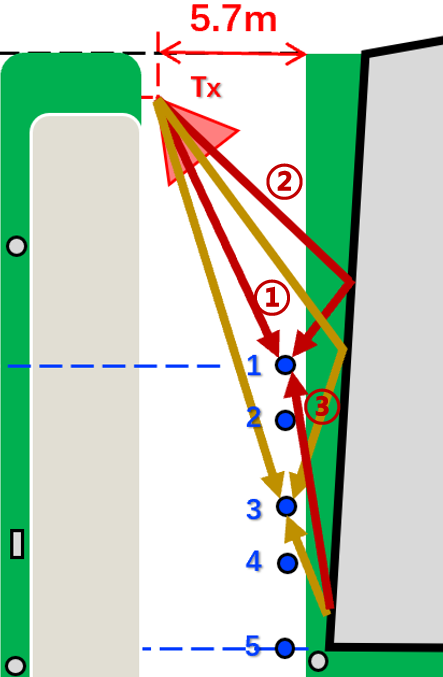}
    }
    \caption{PDAPs and multi-path propagation analysis at Rx1-4.}
    \label{fig:Rx1-4}
\end{figure*}

\begin{figure*}
    \centering
    \subfigure[PDAP at Rx5.]{
    \includegraphics[width=0.37\linewidth]{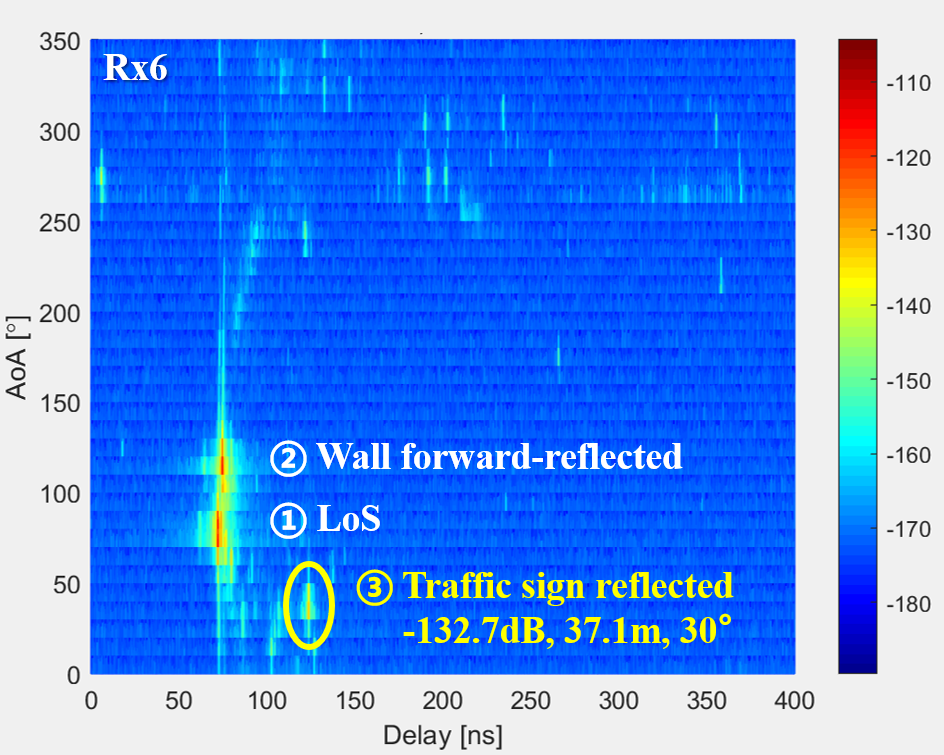}
    }
    \subfigure[PDAP at Rx6.]{
    \includegraphics[width=0.37\linewidth]{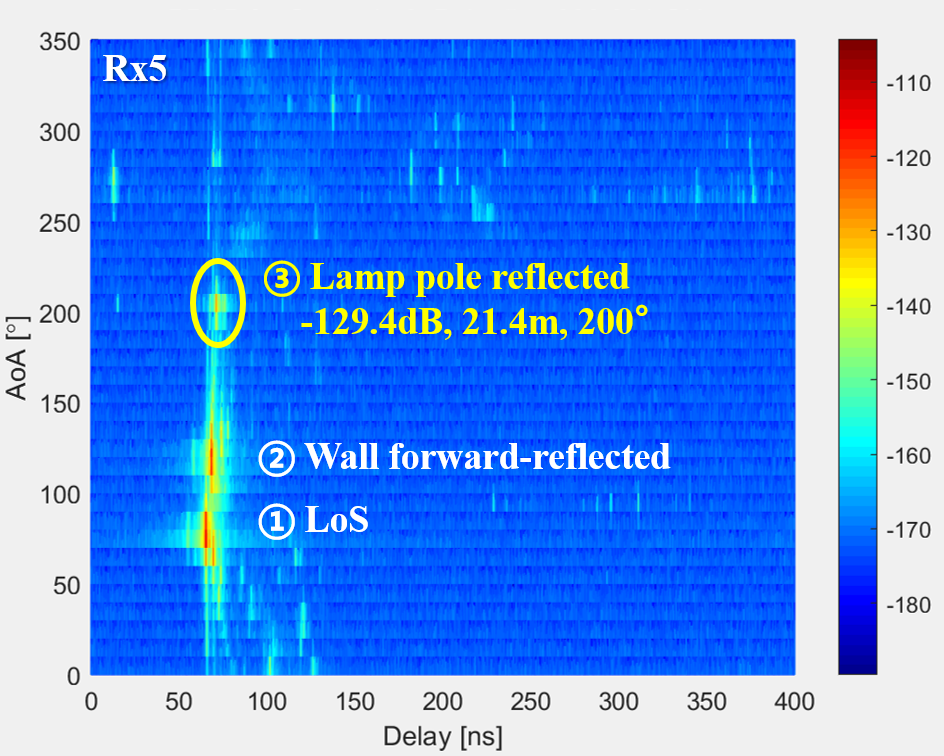}
    }
    \subfigure[Multi-path at Rx5-6.]{
    \includegraphics[width=0.17\linewidth]{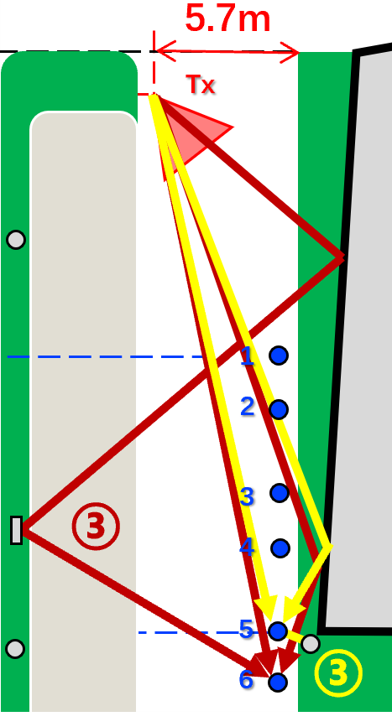}
    }
    \caption{PDAPs and multi-path propagation analysis at Rx5-6.}
    \label{fig:Rx5-6}
\end{figure*}
The PDAP results at 306-321~GHz are taken as examples, while the results in the two bands are summarized and compared in Table~\ref{tab:parameters}.
First, for Rx1 to Rx4, as shown in Fig.~\ref{fig:Rx1-4}, three categories of MPCs are detected, namely, LoS, forward-reflected, and back-reflected rays from the wall. First, the LoS ray with azimuth angle of arrival (AoA, $\varphi$) equals to 60-70$^\circ$, travels 34.27~ns, 40.53~ns, 49.87~ns, and 55.93~ns, to reach Rx1-4. These LoS time-of-arrival (ToAs) correspond to the path length of 10.28~m, 12.16~m, 14.96~m, and 16.78~m, respectively. The power value varies from -104~dB to -113~dB. 

Second, the forward-reflected rays are attributed to the reflection on the west wall of Wenbo Building, with $\varphi$ ranges over 120-130$^\circ$. The traveling path length is around 3~m longer than the LoS ray. The extra loss of wall forward-reflection is measured about 5-15~dB. By contrast, the third type of rays back-reflect from the west wall of Wenbo Building, with $\varphi$ ranges over 250-260$^\circ$, and propagate over 30.6~m to 23.7~m. The power is around -140~dB, which denotes that the extra loss of wall back-reflection is about 20-30~dB, i.e., noticeably higher than the forward-reflection. In our measurement, due to the geometry, the back-reflected rays from Wenbo Building can only be observed at Rx1-4.

Interestingly, between the two wall reflected rays, we observe a tail in PDAP with an increasing delay, a larger AoA and decreasing power, which demonstrates the continuous scattering from the wall. This phenomena occurs since the Tx and Rx positions are close to the wall of Wenbo Building. Therefore, the power of scattered rays from the wall material is above the noise floor. Moreover, as Rx1-4 are distributed in parallel with the wall, we can also observe continuous change of delays (and thus the values of power) and AoAs of scattering rays among PDAPs.

As shown in Fig.~\ref{fig:Rx5-6}, besides the LoS ray and the forward-reflected ray from the wall, reflected rays from the lamp pole and the traffic sign are detected at Rx5 and Rx6. Reflected from the lamp pole at the southwest corner of Wenbo Building, the reflected ray is observed at Rx5 with $\varphi$ equals to 200$^\circ$. 
Moreover, coming from the traffic sign beside the parking lot, the reflected ray is observed at Rx6 with $\varphi$ equals to 30$^\circ$. To compare the reflection losses, the lamp pole reflection loss is about 15~dB, while the traffic sign, as a flat metal plane, results in much lower loss, e.g., around 6-8~dB.

\begin{figure*}
    \centering
    \subfigure[PDAP at Rx8.]{
    \includegraphics[width=0.35\linewidth]{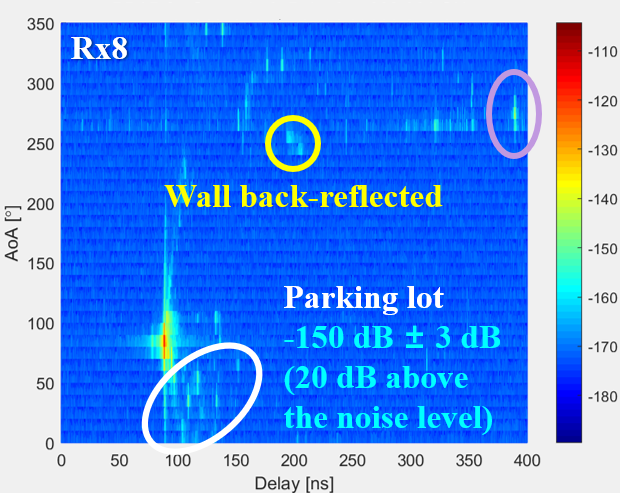}
    }
    \subfigure[PDAP at Rx10.]{
    \includegraphics[width=0.35\linewidth]{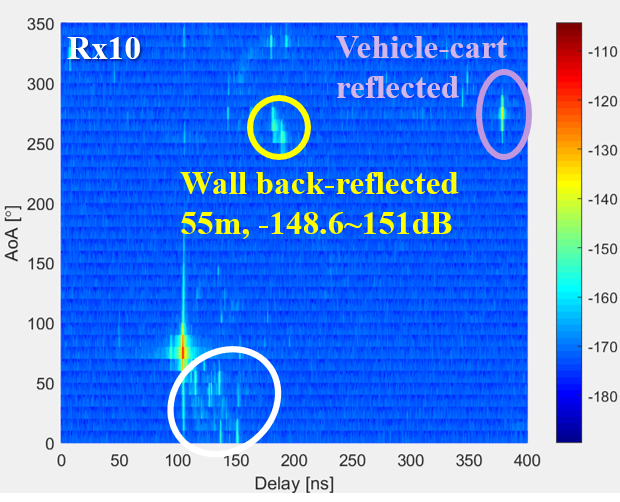}
    }
    \subfigure[The photo of the vehicle.]{
    \includegraphics[width=0.23\linewidth]{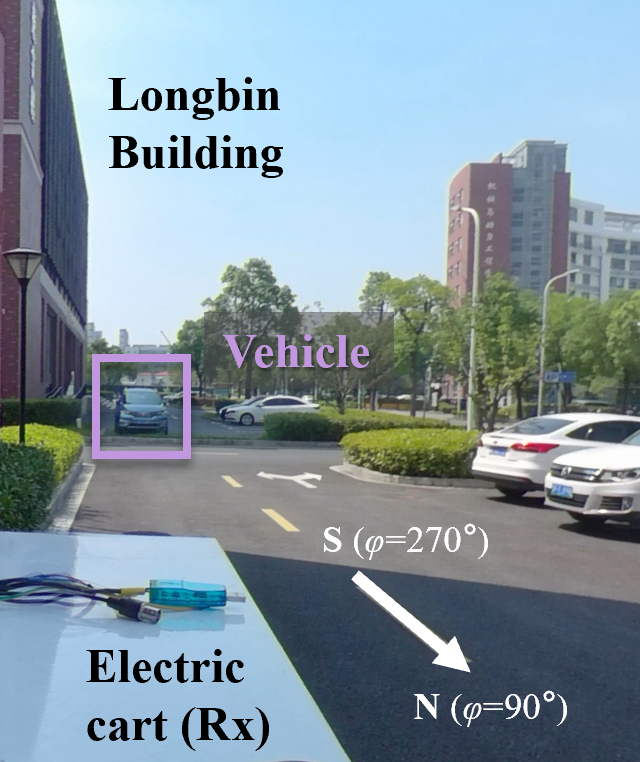}
    }
    \caption{PDAPs at Rx8, Rx10 and the vehicle reflection in the south.}
    \label{fig:Rx7-10}
\end{figure*}

Moving to the south direction further, Rx7-10 are the last four LoS positions, where the received MPCs are similar. Fig.~\ref{fig:Rx7-10} takes Rx8 and Rx10 as examples. First, the LoS ray and the back-reflected rays from Longbin Building can be observed. The back-reflection of the wall causes additional loss of 28-30~dB. Similar to other LoS Rx positions, sporadic rays from the parking lot are detected, i.e., with $\varphi$ in the range of 0-90$^\circ$. The scatterers include vehicles, lamp poles (30$^\circ$- 40$^\circ$), traffic signs (50$^\circ$- 60$^\circ$), and plants. However, unlike strong reflected rays (e.g., with power around -130~dB) at Rx5 and Rx6, these scattering rays are much weaker, with power around -150~dB, i.e., additional 20~dB losses.

%\begin{figure}
%    \centering
%    \includegraphics[width=\linewidth]{figures/strip250-270.png}
%    \caption{PDAP strips around 260$^\circ$ at Rx1-8.}
%    \label{fig:strip}
%\end{figure}
%Furthermore, we observe successive rays around 270$^\circ$ at Rx7 and Rx8, but not at Rx9 or Rx10. The reason is that Rx7 and Rx8 are located on the north-south street, while Rx9 and Rx10 are on the other street. For all points on the north-south street, Rx1 to Rx8, the same phenomena exist. As depicted in Fig.~\ref{fig:strip}, the rays occur at the same angle around 260$^\circ$ and 270$^\circ$ (rarely at 250$^\circ$), and travels shorter distance (i.e., with smaller delay) to reach the receiver as the Rx moves from Rx1 to Rx8, which indicates that the scatterer is in the south. According to the photos that we take at that day as Fig.~\ref{fig:Rx7-10}(c), the vehicle parked near Longbin Building is likely the scatterer, and the minimal reflection loss is 16~dB. Note that the delay value is recorded cyclically with the duration of 400~ns, and therefore, MPCs with delay smaller than 50~ns actually travels 400-450~ns.

\begin{figure}
    \centering
    \subfigure[Multi-path at Rx11-16.]{
    \includegraphics[width=0.43\linewidth]{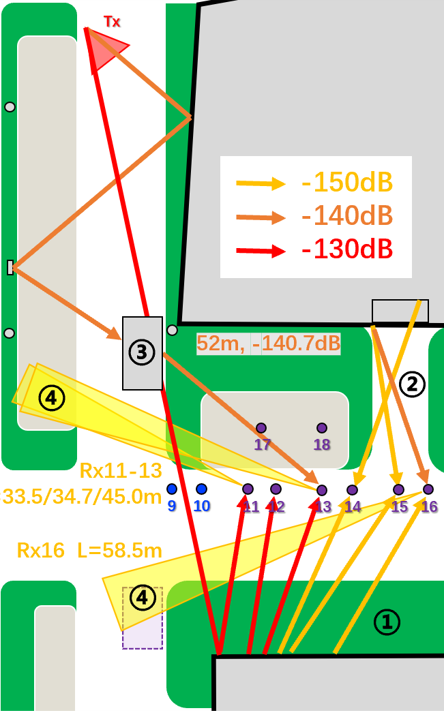}
    }
    \subfigure[Multi-path at Rx17-18.]{
    \includegraphics[width=0.43\linewidth]{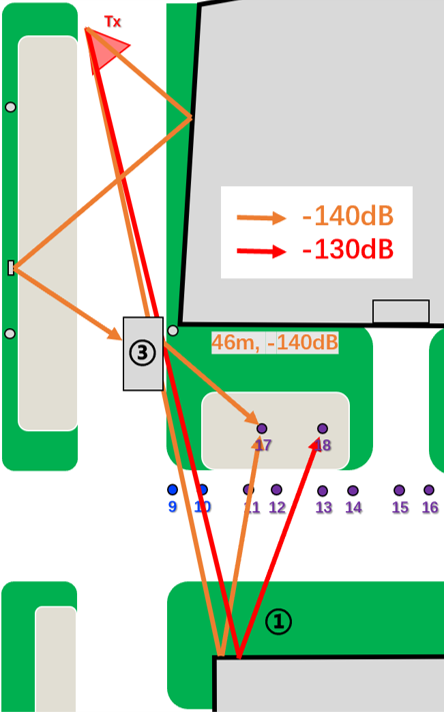}
    }
    \caption{Multi-path analysis at Rx11-18 (NLoS Rx points).}
    \label{fig:Rx11-18}
\end{figure}

When the receiver moves to NLoS positions, as noted in Section~\ref{section: campaign}, a hybrid vehicle is parked at the southwest corner of Wenbo Building to outsource power for the channel sounder system. As summarized in Fig.~\ref{fig:Rx11-18}, the rays mainly come from (i) the first-order forward-reflection from Longbin Building, (ii) the higher-order reflection between two buildings, (iii) the transmission through the vehicle, and (iv) the scattering from vehicles in the parking area and near Longbin Building. First, the highest received power is -130~dB in the NLoS case. Second, the reflection loss caused by the wall of Longbin Building varies from 18~dB to 30~dB, depending on the angle of incidence. Third, the transmission loss through the vehicle is less than 10~dB, deduced 
based on the comparison with the reflected ray from the traffic sign at Rx6. By contrast, the power of the scattering rays from vehicles is as low as -150~dB at NLoS Rx points, which indicates less significance of scattering (compared to reflection) on vehicles.

\subsection{Characteristic Analysis}
Based on the data post-processing procedures as detailed in~\cite{li2022modifiedSAGE}, we present the channel characteristics in depth. We clarify that we only focus on MPCs within the dynamic range of 30~dB, in order to be compatible with the lower dynamic range of general THz devices~\cite{channel_tutorial}.
In Table~\ref{tab:parameters}, statistical channel parameters in two frequency bands are summarized and compared with those in the UMi-street canyon scenario specified in 3GPP TR~38.901~\cite{3gpp38901}, which are discussed as follows.

\subsubsection{Path loss and shadow fading}
The best direction path loss and the omni-directional path loss~\cite{wang2022thz} are investigated.
The close-in (CI) and $\alpha$-$\beta$ path loss models are invoked, which are expressed respectively as
\begin{subequations}
\begin{align}
\text{PL}^{\rm CI} &= 10\times\text{PLE}\times\log_{10}\frac{d}{d_0}+\text{FSPL}(d_0)+X^{\rm CI}_{\sigma_{\rm SF}},\\
\text{PL}^{\alpha\beta} &= 10\times\alpha\times\log_{10}d+\beta+X^{\alpha\beta}_{\sigma_{\rm SF}},
\end{align}
\end{subequations}
where PLE is the path loss exponent and $d$ denotes the distance between Tx and Rx. The reference distance $d_0$ is equal to 1~m. The free-space path loss (FSPL) is given by the Friis’ law. Moreover, $\alpha$ is the slope coefficient and $\beta$ represents the optimized path loss offset in dB. $X^{\rm CI}_{\sigma_{\rm SF}}$ and $X^{\rm \alpha\beta}_{\sigma_{\rm SF}}$ are zero-mean Gaussian random variables with standard deviation $\sigma^{\rm CI}_{\rm SF}$ and $\sigma^{\alpha\beta}_{\rm SF}$ in dB, respectively, indicating the fluctuation caused by shadow fading.

The observations are drawn as follows.
First, the omni-directional PLEs are smaller since the power of all MPCs are added up, resulting in a lower omni-directional path loss.
Second, compared with the UMi-street canyon scenario specified in 3GPP TR~38.901, the omni-directional PLE in the THz band is smaller in the LoS case but larger in the NLoS case.
Third, compared to large-scale characteristics at 306-321~GHz, the path losses at 356-371~GHz are higher by 4-6~dB, and the shadow fading effect at 356-371~GHz is more significant, as expected.

\subsubsection{Cluster parameters}
The numbers of clusters in the outdoor L-shaped street in the THz band are much smaller than that is specified in 3GPP UMi-street canyon scenario in frequency bands lower than 100~GHz. This fact indicates the strong sparsity in THz bands. To be specific,  the average number of clusters at 300~GHz and 350~GHz is about 2-3 in LoS cases and 5-6 in NLoS cases. Furthermore, due to the sparsity of MPCs under the target 30~dB dynamic range, MPCs inside each cluster are more confined, and thus the intra-cluster delay and angular spreads are also much smaller.

\subsubsection{Delay and angular spreads}
We use the root-mean-square (RMS) delay spread (DS), azimuth spread of angle (ASA) and elevation spread of angle (ESA) to measure the power dispersion of MPCs in temporal and spatial domains~\cite{wang2022thz}.
First, the values of DS, ASA, and ESA in the NLoS case are noticeably larger than the counterparts in the LoS case.
Second, in the LoS case, DS and ASA are normally smaller than 10~ns and 20$^\circ$. For Rx points where the long vehicle-reflected ray is detected, the value of DS increases, which is still smaller than 20~ns at 306-321~GHz.
Furthermore, the higher frequency band over 356-371 GHz has smaller DS and ASA.
Besides, the values of spreads in the L-shaped street in the THz band are smaller than those in the UMi-street canyon scenario in 3GPP TR~38.901 in frequency bands lower than 100~GHz. Compared to results in the indoor L-shaped hallway\footnote{The preliminary results in the indoor L-shaped hallway are presented in~\cite{wang2022thz}. The measurement campaign is later extended to both 306-321~GHz and 356-371~GHz, and the complete characteristic analysis results are updated according to the same data processing procedure as this paper.}, DS in the outdoor L-shaped street is smaller in the LoS case, while DS values in the NLoS case are comparable, in both frequency bands. ASAs are smaller in the outdoor L-shaped street for both cases and both frequency bands. Smaller power spreads in the outdoor L-shaped street in both temporal and spatial domains are attributed to the absence of walls that surround the scenario and thus confine MPCs with diverse delays and AoAs to the scenario.

\begin{table}
\caption{Summary of statistical channel parameters.}
\centering
\includegraphics[width=\linewidth]{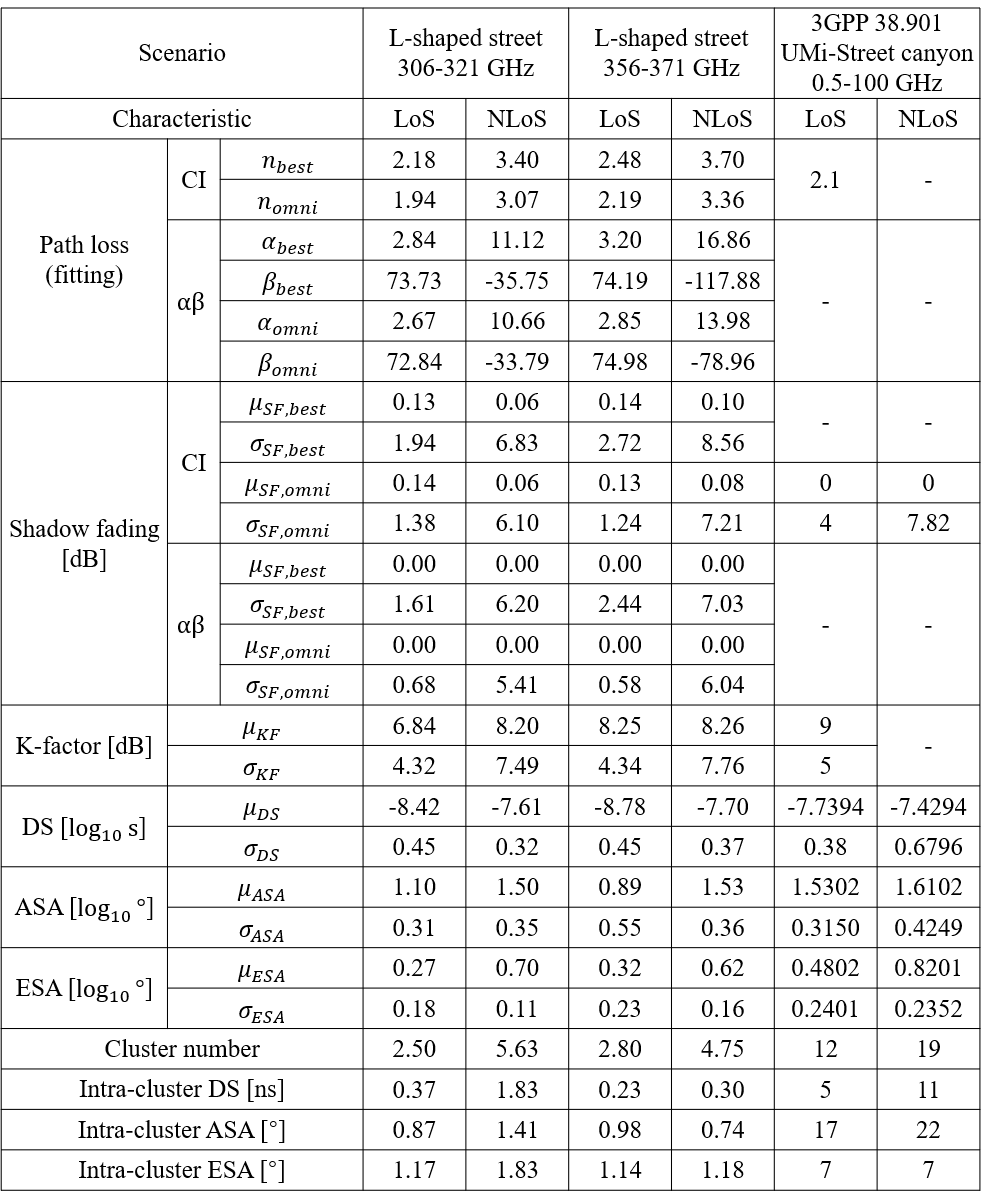}
\label{tab:parameters}
\end{table}

%\subsubsection{Comparison between L-shaped indoor and outdoor scenarios}

\section{Conclusion} \label{section: conclusion}
In this paper, we conducted a wideband channel measurement campaign in an L-shaped street on the university campus at 306-321~GHz and 356-371~GHz.
The scenario contains one street along the parking area, and another perpendicular street canyon between two buildings.
The characteristics of multi-path propagation are elaborated to identify major scatterers in the environment and their impact on MPCs. We analyze the outdoor THz channel characteristics in the two frequency bands, and draw the following observations.

\begin{itemize}
    \item First, reflected and scattered rays on the wall are well observed. Forward-reflected rays endure extra loss about 5-15~dB, while back-reflected rays endure extra loss about 20-30~dB. Besides, the abundance of scattering on the wall is demonstrated by PDAP results.
    \item Second, other major scatterers, i.e., traffic signs, lamps, and vehicles, are identified in this outdoor scenario. For instance, the traffic sign and the lamp result in additional reflection losses about several decibels and 15~dB, respectively. 
    \item Third, the interaction on the vehicle is investigated. While transmission or reflection renders an extra loss less than 10~dB, the scattering on vehicles are much less significant.
    \item Fourth, as the receiving point moves from LoS positions into the street canyon (Street B) between two high buildings, the LoS is blocked by the building. For NLoS receiving points as far as 20~m away from the open crossroad, i.e., the entrance of the street canyon, the received power is 20~dB lower than that of the LoS point at the crossroad.
    \item Fifth, we further analyzed THz channel characteristics in the outdoor scenario, and compared our results with those in the UMi-street canyon scenario below 100~GHz, which indicates the sparsity of MPCs and smaller power spreads in both temporal and spatial domains in the THz band.
\end{itemize}

%including path losses, shadow fading, cluster parameters, delay spread and angular spread, and compared them with the counterparts of the street canyon scenario in lower frequency bands, as summarized in Table~\ref{tab:parameters}.

%Acknowledgement: Chong Han acknowledges the support from the National Natural Science Foundation of China (NSFC) under Grant 62027806.

%\balance

\bibliographystyle{IEEEtran}
\bibliography{bibliography}

\end{document}